# Electronic transport and Fermi surface of Weyl semimetal WTe$_2$: quantum oscillations and first-principles study

B.M. Fominykh[1,*], A.N. Perevalova[1], S.T. Baidak[1], A.V. Lukoyanov[1,2], S.V. Naumov[1], E.B. Marchenkova[1], V.V. Marchenkov[1,2]

[1] M.N. Mikheev Institute of Metal Physics, Ural Branch of the Russian Academy of Sciences, 620108, S. Kovalevskaya str., 18, Ekaterinburg, Russia

[2] Institute of Physics and Technology, Ural Federal University, 620002, Mira str., 21, Ekaterinburg, Russia

E-mail: bogdan.fominyh@mail.ru, domozhirova@imp.uran.ru, march@imp.uran.ru

* Corresponding author

**Abstract.** Currently, topological semimetals are being actively investigated from both theoretical and experimental perspectives due to their unique physical properties, including topologically protected states, large magnetoresistivity, and high carrier mobility, which make these materials promising for various applications in electronics. In this work, we present experimental and theoretical studies of the electronic structure and electronic transport in the Weyl semimetal WTe$_2$. Band structure of WTe$_2$ was scrutinized with DFT+U+SOC method showing the semimetallic nature and sensitivity of the structure to the value of U and to changes in the Fermi energy. Our results demonstrate that WTe$_2$ is in a near-compensated state and exhibits an almost quadratic non-saturating magnetoresistivity. It is found that WTe$_2$ violates the classical Kohler's rule, which is attributed to the coexistence of multiple scattering mechanisms and a strong temperature dependence of the current carrier concentration. Analysis of the Shubnikov–de Haas oscillations reveals three distinct frequencies corresponding to two electron and one hole Fermi surface pockets, which are well reproduced in Fermi surface calculations. Using the Lifshitz-Kosevich formalism, we determined the electronic structure parameters for each Fermi surface pocket. Additionally, we discuss the relationship between the g-factor and the Berry phase extracted from quantum oscillations.

**Keywords**: topological semimetals, Kohler's rule, quantum oscillations, Berry phase, Fermi surface, DFT calculations.

## 1. Introduction

Transition metal dichalcogenides occupy a one of the central place in modern condensed matter physics due to their unique electronic properties, including superconductivity, charge density waves, and nontrivial topology. Unlike topological insulators, which possess a bulk bandgap and conducting surface states, topological semimetals are characterized by special points or lines in the bulk band structure where the valence and conduction bands cross each other linearly. These materials, such as Dirac and Weyl semimetals, exhibit exotic electronic properties, including linear dispersion, Fermi arcs, and high charge carrier mobility. Among them, tungsten ditelluride stands out as a prototypical compensated semimetal, demonstrating giant non-saturating magnetoresistivity (MR) [1], pressure-induced superconductivity [2], and predicted topological states [3-5]. Topological semimetals hold promise for modern technologies, including superconducting devices, highly sensitive photodetectors and terahertz radiation sensors, as well as low-power spintronic applications [6-13]. These materials open new avenues for fundamental research and practical applications, making them one of the most promising frontiers in condensed matter physics [14, 15]. However, while their exotic transport properties stem from a complex Fermi surface with nearly equal electron and hole concentrations, questions regarding the precise electronic structure remain debated.

A key unresolved issue is the origin of quantum oscillations in $WTe_2$: whether they arise from trivial or topologically nontrivial Fermi surface components. Despite the widespread use of Shubnikov–de Haas (SdH) oscillations for topological studies [16-19], data interpretation is complicated by multiple frequencies, beating patterns, and conflicting reports on effective masses and Berry phases. The scatter in reported oscillation frequencies, carrier mobilities, and scattering mechanisms further highlights the sensitivity of electronic transport properties of $WTe_2$ to sample quality and measurement conditions.

This study bridges high-precision magnetotransport experiments with first-principles calculations to unravel the intricate electronic properties of $WTe_2$. By focusing on high-quality single crystals, we address key challenges in the field, including the interpretation of multi-frequency quantum oscillations and the role of Berry phase in topological materials. Our work goes beyond conventional analyses by critically examining the limitations of quantum oscillation techniques, particularly in systems with large g-factors and complex Fermi surfaces. Previous studies primarily focused on determining oscillation frequencies and effective masses, comparing them with first-principles calculations. In our work, we employ the multi-band Lifshitz-Kosevich (LK) formalism to extract all possible transport and electronic structure parameters, including the Dingle temperature and Berry phase. Moreover, our comparative analysis revealed a discrepancy: while our calculations suggest a non-trivial Berry phase for hole pockets, the LK formula analysis indicates a non-trivial Berry phase for electron pockets. We attribute this inconsistency to the potential Zeeman effect arising from a large g-factor, a particularly relevant consideration for topological semimetals with strong spin-orbit coupling. Notably, prior works often neglect the g-factor. Additionally, we applied the extended Kohler's rule proposed in [20] to this material for the first time, observing strong agreement between the relative carrier concentrations derived from extended Kohler's rule and those obtained from the two-band model. Our results not only refine the electronic structure of $WTe_2$ but also provide a methodology for studying quantum oscillations in multiband topological materials.

**2. Materials and methods**

Single-crystal $WTe_2$ was grown by chemical vapor transport, following the procedure described in Ref. [21]. The crystal structure was verified by X-ray diffraction (XRD) on a DRON-2.0 diffractometer (Cr Kα radiation). Figure 1(a) shows a fragment of the XRD pattern obtained from a ground $WTe_2$ crystal. The grown single crystal exhibits an orthorhombic lattice with the space group *Pmn*$2_1$ (No. 31) and lattice parameters a = 3.435(8) Å, b =

6.312(7) Å, and c = 14.070(4) Å. The crystal structure of WTe$_2$, schematically illustrated in the inset of Fig. 1(a), consists of triple atomic layers (Te–W–Te) bound by weak van der Waals interactions, while strong covalent bonding dominates within each layer. These layers stack along the c-axis, with the unit cell containing four formula units. Surface microstructure and chemical composition were analyzed using a Tescan MIRA LMS scanning electron microscope (SEM) equipped with energy-dispersive X-ray spectroscopy (EDS) at the Collaborative Access Center "Testing Center of Nanotechnology and Advanced Materials" of M. N. Mikheev Institute of Metal Physics of the Ural Branch of the Russian Academy of Sciences (IMP UB RAS). The layered structure of the WTe$_2$ single crystal is evident in the SEM image shown in Fig. 1(b). EDS analysis confirmed a near-stoichiometric composition, with atomic concentrations of W (34.53%) and Te (65.47%).

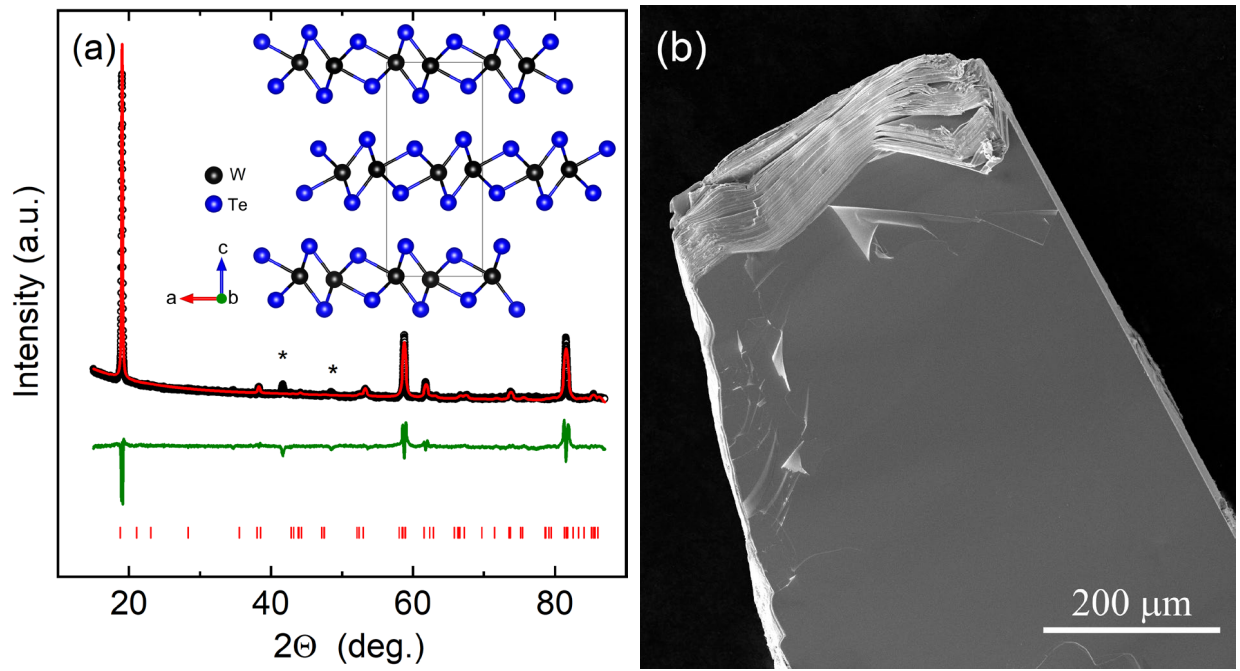

Fig. 1. (a) Fragment of the XRD pattern obtained from a ground WTe$_2$ single crystal. Symbols represent experimental data, while the solid red line shows the Rietveld refinement. Vertical tick marks at the bottom indicate Bragg peak positions for the *Pmn*2$_1$ structure. Asterisks denote traces of tellurium impurities. Inset: Schematic of the WTe$_2$ crystal structure, where black and blue circles represent W and Te atoms, respectively. Arrows indicate crystallographic axes directions. (b) SEM image of the WTe$_2$ single crystal surface microstructure acquired in secondary electron (SE) contrast mode.

For transport measurements, a 40 µm-thick flake was mechanically exfoliated from the as-grown single crystal. Resistivity, magnetoresistivity, and Hall effect measurements were performed using a standard four-probe configuration in the temperature range of 2-300 K under magnetic fields up to 9 T. The magnetic field was applied along the crystallographic c-axis, while the electric current flowed within the ab-plane of the crystal. All transport measurements were conducted using a PPMS-9 system at the Collaborative Access Center "Testing Center of Nanotechnology and Advanced Materials" of the IMP UB RAS.

Our electronic structure calculations in this work were conducted within density functional theory so-called first-principles method DFT+U+SOC [22]. The basic DFT functional was chosen to be generalized gradient approximation (GGA) in the version of Perdew-Burke-Ernzerhof (PBE). The ultrasoft full relativistic pseudopotentials were taken to account for spin-orbit coupling (SOC) in the considered $WTe_2$ compound, this approach is known to provide reliable electronic structure and other physical characteristics. The calculations were done using the Quantum ESPRESSO software package [23, 24], and the pseudopotentials were chosen from its libraries. The wave functions were presented in the basis of plane waves. The reciprocal space was divided into a grid of 12 x 12 x 12 k-points which is enough for good convergence of the results. Kinetic energy cutoffs for the wavefunctions and for the charge density were chosen as 70 Ry and 700 Ry respectively which was found to be enough for the self-consistency of the results. When constructing densities of states, the value of the gaussian broadening was chosen to be 0.01 Ry for the density peaks to be looking smoother. To calculate the value of U for the W 5d states in $WTe_2$ with our sample-characteristic crystal structure parameters, hp.x code [25] in Quantum ESPRESSO was employed that resulted in 3.0 eV. Fermi surface sections and positions of the Weyl nodes were calculated in the basis of Wannier functions with WannierTools software package [26].

## 3. Results and discussion

### 3.1 Electronic transport properties

#### 3.1.1 Electrical resistivity and galvanomagnetic properties

Figure 2 shows the temperature dependence of the electrical resistivity of $WTe_2$ in zero magnetic field and under a 9 T field. The residual resistivity ratio (RRR = $\rho_{300 K}/\rho_{2 K}$) for the studied single crystal is 38. The $\rho(T)$ dependence exhibits metallic behavior throughout the entire temperature range. In the low-temperature region (below 70 K), the resistivity follows a $T^2$ dependence, which originates from both electron-electron scattering and the interference mechanism of "electron-phonon-surface" scattering [21, 27, 28]. At higher temperatures, the $T^2$ contribution is suppressed and the dependence becomes nearly linear due to dominant electron-phonon scattering. The applied magnetic field B leads to an increase in the resistivity and induces a minimum in the $\rho(T)$ dependence at 60 K. As demonstrated in our previous works [21, 27, 28], this minimum corresponds to the crossover between effectively high ($\omega_c\tau \gg 1$) and weak ($\omega_c\tau \ll 1$) magnetic field regimes, where $\omega_c$ is the cyclotron frequency and $\tau$ is the relaxation time. Qualitatively, this can be understood as follows. In compensated conductors, when $\omega_c\tau \gg 1$, electrons complete many cyclotron orbits between scattering events, leading to an increase in resistivity due to the restricted transverse motion of charge carriers in the magnetic field. In contrast, when $\omega_c\tau \ll 1$, electrons undergo collisions before completing a full cyclotron orbit, weakening the influence of the field. In this regime, resistivity is primarily governed by phonon and defect scattering. The inset to Fig. 2 displays the temperature dependence of the magnetoresistivity (MR = $[(\rho_{xx} - \rho_0)/\rho_0]\times 100\%$, where $\rho_{xx}$ is the resistivity in a magnetic field B, $\rho_0$ is the electrical resistivity without a field), which decreases rapidly from 1900% at 2 K to 1.5% at 300 K, in agreement with other reported studies [29].

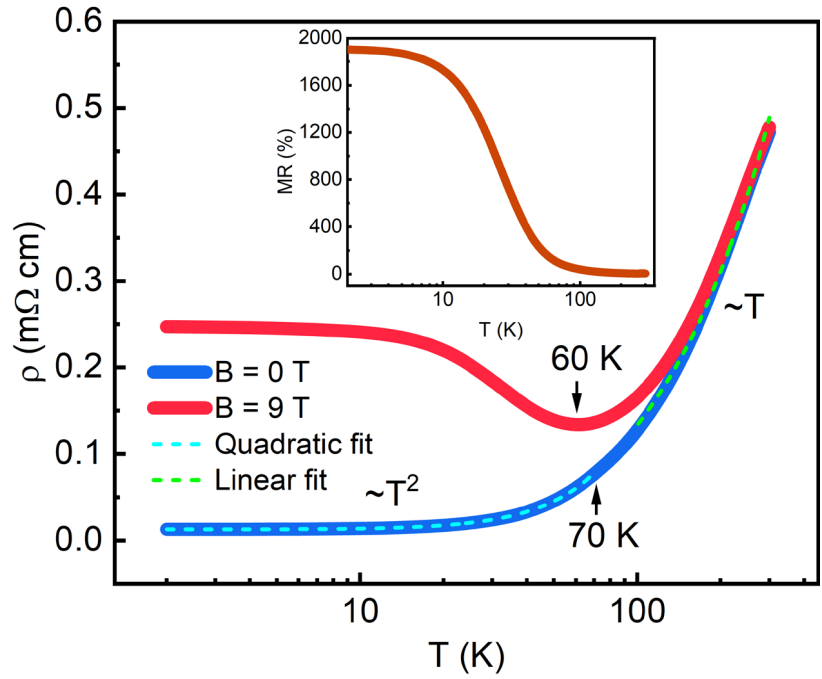

Fig. 2. Temperature dependence of the electrical resistivity of WTe$_2$ in zero magnetic field and under a 9 T magnetic field. The inset shows the temperature dependence of MR at 9 T.

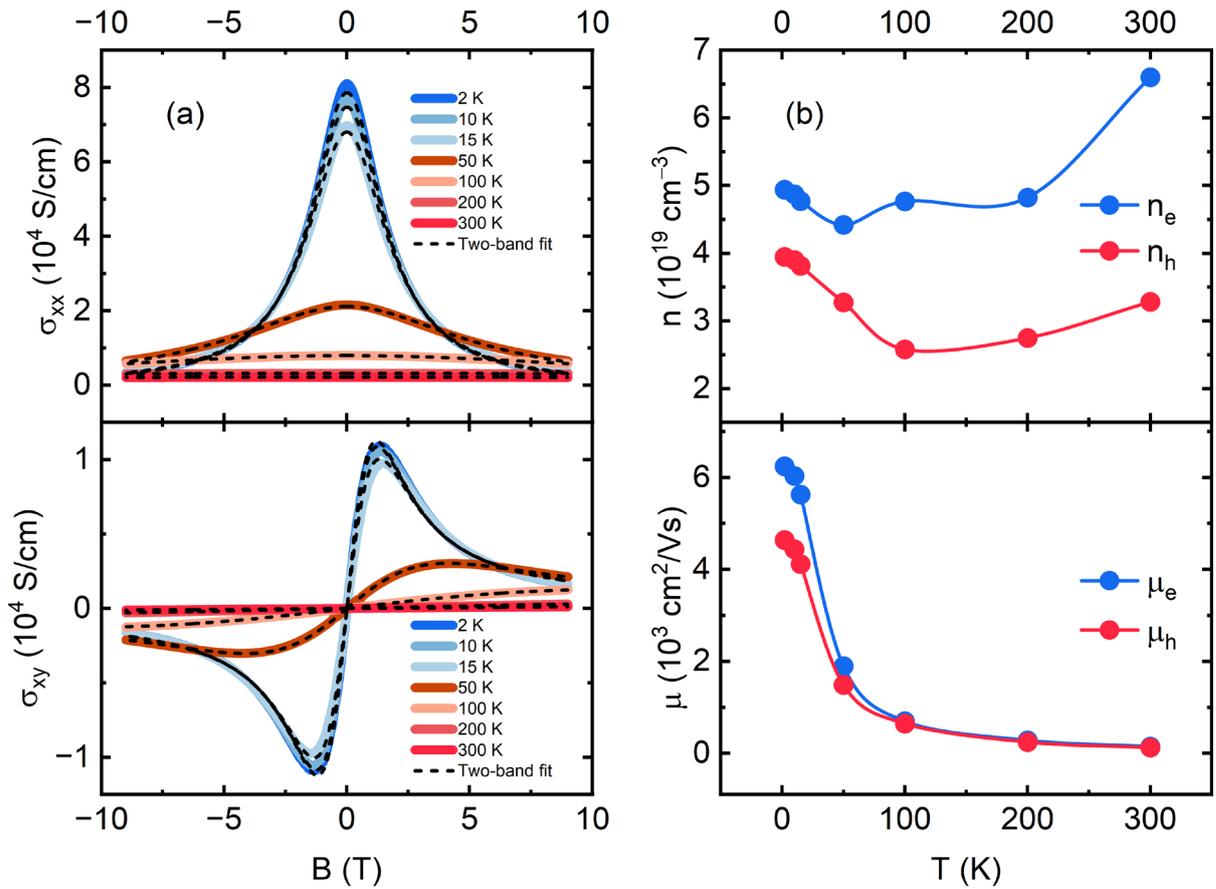

Fig. 3. (a) Field-dependent longitudinal conductivity ($\sigma_{xx}$) and Hall conductivity ($\sigma_{xy}$) of the WTe$_2$ single crystal at various temperatures in magnetic fields up to 9 T. Dashed lines show fits using the two-band model. (b) Temperature evolution of carrier concentrations and mobilities obtained from fitting the $\sigma_{xx}(B)$ and $\sigma_{xy}(B)$ data with Eqs. (1) and (2).

The two-band model was employed to evaluate charge carrier concentrations and mobilities in $WTe_2$, which has been successfully used to describe transport properties of various topological semimetals [30, 31]. From the experimental data of $\rho_{xx}$ and $\rho_{xy}$, we calculated the conductivity tensor components: the conductivity in a magnetic field $\sigma_{xx} = \frac{\rho_{xx}}{\rho_{xx}^2 + \rho_{xy}^2}$ and the Hall conductivity $\sigma_{xy} = -\frac{\rho_{xy}}{\rho_{xy}^2 + \rho_{xx}^2}$. The field dependences of $\sigma_{xx}(B)$ and $\sigma_{xy}(B)$ were fitted within the two-band model using the following equations:

$$\sigma_{xx}(B) = \frac{en_e\mu_e}{1+(\mu_e B)^2} + \frac{en_h\mu_h}{1+(\mu_h B)^2}, \tag{1}$$

$$\sigma_{xy}(B) = \frac{en_e\mu_e^2 B}{1+(\mu_e B)^2} - \frac{en_h\mu_h^2 B}{1+(\mu_h B)^2}, \tag{2}$$

where $n_e$ ($n_h$) and $\mu_e$ ($\mu_h$) represent the electron (hole) concentrations and mobilities, respectively. The dependences of $\sigma_{xx}(B)$ and $\sigma_{xy}(B)$ at various temperatures from 2 to 300 K are shown in Fig. 3(a). Dashed lines indicate the fitting curves obtained from the two-band model, which agree well with the experimental data. The electron and hole concentrations and mobilities calculated as fitting parameters are presented in Fig. 3(b) over the temperature range from 2 to 300 K. At T = 2 K, the values are $n_e$ = 4.9×10$^{19}$ cm$^{-3}$, $n_h$ = 3.9×10$^{19}$ cm$^{-3}$, $\mu_e$ = 6.2×10$^3$ cm$^2$/(V·s), and $\mu_h$ = 4.6×10$^3$ cm$^2$/(V·s). Notably, the electron and hole concentrations are almost equal below 50 K, indicating that $WTe_2$ is a nearly compensated semimetal. Both electron and hole mobilities decrease with increasing temperature due to enhanced carrier scattering.

Fig. 4 shows the field dependence of MR, reaching 1900% at 2 K in a field of 9 T. The MR increases monotonically without showing any signs of saturation, with a nearly quadratic field dependence. Such non-saturating ~$B^2$ behavior is characteristic of compensated conductors with closed Fermi surfaces [32]. Similar behavior has been reported in other studies [1, 3, 30, 33-38].

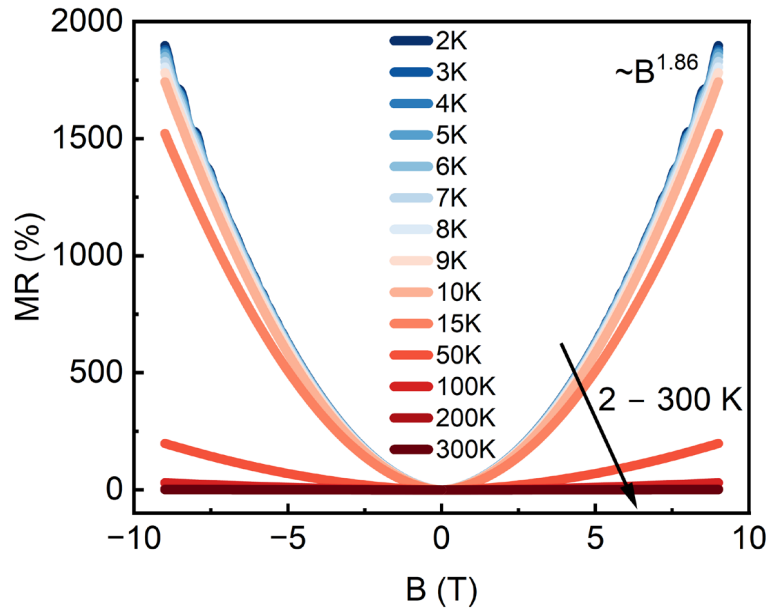

Fig. 4. Magnetic field dependence of the MR of $WTe_2$ at different temperatures.

## 3.1.2 Kohler's rule

Magnetic field dependences of the MR of WTe$_2$ in logarithmic scale at different temperatures are shown in Fig 5(a). It is well established that in metallic conductors, MR should obey Kohler's rule. According to this rule, if transport properties are dominated by a single scattering mechanism, the MR measured at different temperatures should collapse onto a universal curve MR = f(B/ρ$_0$) [20]. The Kohler plot shown in Fig. 5(b) reveals that the MR data do not follow the single universal curve, particularly at higher temperatures. This indicates the presence of multiple scattering mechanisms in our sample, consistent with our analysis of temperature-dependent resistivity.

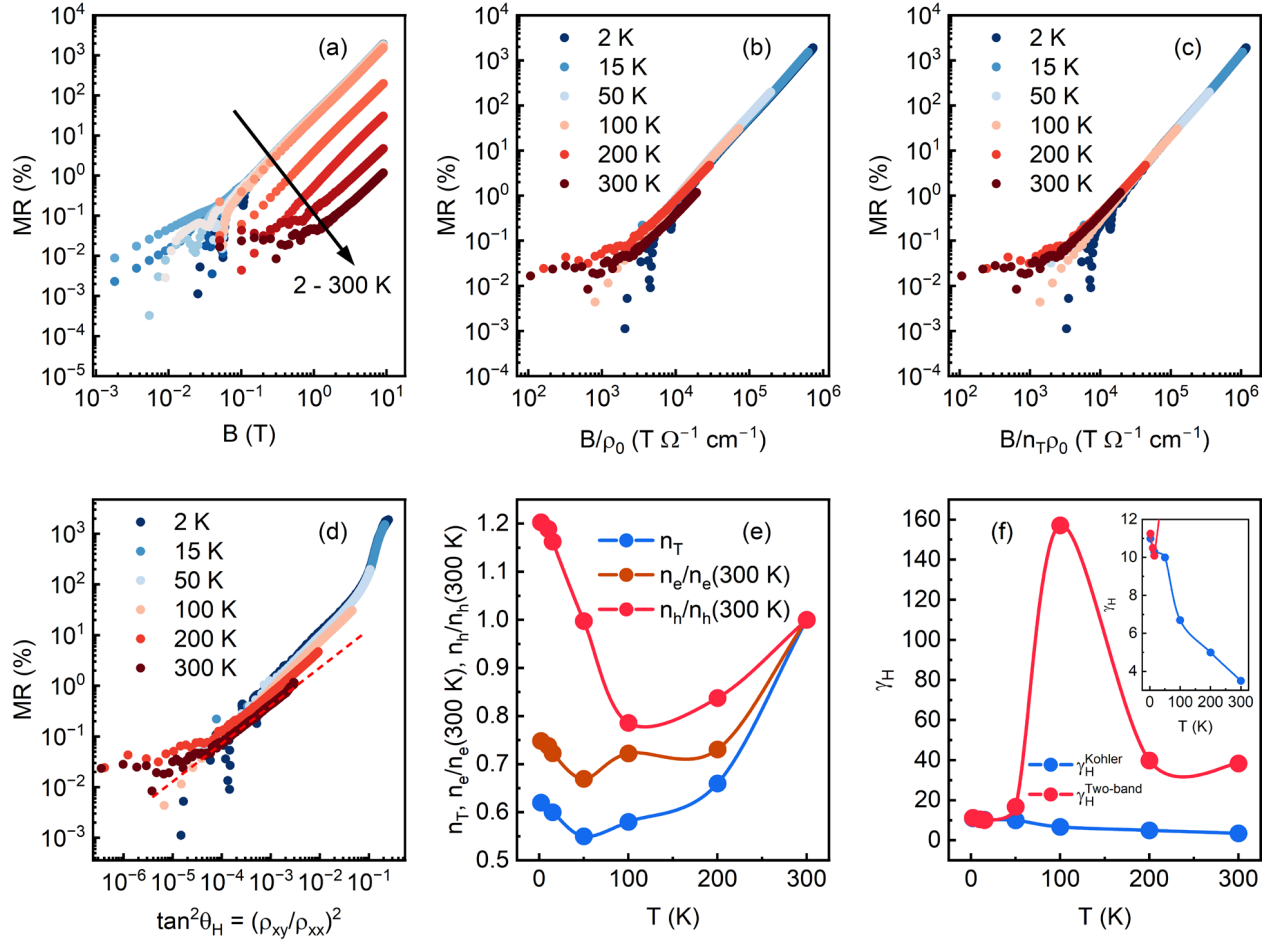

Fig. 5. (a) Magnetic field dependence of the MR of WTe$_2$ in logarithmic scale at different temperatures. (b) Kohler plot MR = f(B/ρ$_0$) and (c) extended Kohler plot MR = f(B/n$_T$ρ$_0$) at different temperatures. (d) MR versus $\tan^2\theta_H$. The dashed straight red line is $MR = \gamma_H \tan^2\theta_H$ with $\gamma_H = 3.5$. (e) Temperature dependence of n$_T$, n$_e$/n$_e$(300 K) and n$_h$/n$_h$(300 K). (f) Temperature dependence of $\gamma_H$, determined from the Kohler plot analysis and the two-band model.

It should be noted that the classical Kohler's rule holds true for most metals where carrier concentration remains nearly temperature-independent. However, in materials such as Weyl semimetals, semiconductors, and superconductors [20, 39, 40], Kohler's rule is frequently violated. This occurs because the charge carrier concentration changes significantly with temperature due to thermal excitation. As demonstrated by our two-band model analysis (Fig. 3), both electron and hole concentrations in WTe$_2$ exhibit strong temperature dependence. To account for this effect, we employed the extended Kohler's rule proposed in Ref. [20]

$$MR = f\left(\frac{B}{n_T \rho_0}\right), \qquad (3)$$

where $n_T$ is a dimensionless parameter describing the relative change in carrier concentration with temperature ($n_T$ = 1 corresponds to classical Kohler's rule). We adjusted $n_T$ to make all curves collapse onto the 300 K curve (where $n_T$ = 1), as shown in Fig. 5(c). The temperature dependence of $n_T$ (Fig. 5(e)) reveals a pronounced non-monotonic behavior that qualitatively agrees well with the normalized electron concentration dependence $n_e/n_e$(300 K) obtained from the two-band model analysis. This agreement suggests that the temperature-induced variation in electron concentration represents one of the key factors responsible for the observed deviations from classical Kohler's rule in WTe$_2$.

In addition to the classical form, Kohler's rule can be expressed in various alternative formulations. One of the most common representations is [20]

$$MR = \gamma_H \tan^2 \theta_H, \qquad (4)$$

where $\theta_H = \arctan(\rho_{xy}/\rho_{xx})$ is the Hall angle and $\gamma_H = (\mu_h/\mu_e)/(1-\mu_h/\mu_e)^2$. From Fig. 5(d), it is evident that in this representation, the MR data still do not collapse onto a single curve. At high fields and low temperatures, MR deviates significantly from the $\tan^2 \theta_H$ proportionality. However, in weak fields, the relation MR ~ $\tan^2 \theta_H$ holds, allowing extraction of the parameter $\gamma_H$. Figure 5(f) shows that the parameter $\gamma_H$, determined from both Kohler plot analysis and the two-band model, exhibits good agreement at low temperatures but shows significant deviation at higher temperatures. This further confirms that the strong violation of Kohler's rule in WTe$_2$ results from both the coexistence of multiple scattering mechanisms and the pronounced temperature dependence of charge carrier concentration.

It should be noted that, at present, there are not many studies devoted to investigating Kohler's rule in WTe$_2$. For example, in [41], the authors studied magnetoresistivity in WTe$_2$ up to 56 T and also observed a violation of Kohler's rule with increasing temperature, similar to our case. The authors of [41] suggest that different types of scattering mechanisms, which affect different types of charge carriers in distinct ways, play a significant role in the measured temperature range. They primarily attribute this to small-angle electron-phonon scattering, which becomes more efficient with increasing temperature. In our work, from the analysis of temperature-dependent resistivity (Fig. 2), we indeed found that as temperature increases, the ρ(T) dependence becomes more linear, indicating enhanced electron-phonon scattering at higher temperatures.

Another interesting result was reported in [42]. In contrast to [41] and our work, the authors of [42] found that the temperature-dependent magnetoresistivity of WTe$_2$ measured at different magnetic fields collapsed onto a universal curve MR = f(B/ρ$_0$), i.e., it obeyed Kohler's rule. At the same time, by analyzing the components of the resistivity tensor ρ$_{xx}$ and ρ$_{xy}$ at different temperatures using a two-band model, the authors of [42] demonstrated that the electron and hole concentrations were much closer to each other and exhibited significantly weaker temperature dependence compared to our results. This observation also correlates with the fact that in [42], the MR followed a nearly quadratic dependence MR ~ $B^{1.92}$, whereas in our case, it deviated slightly MR ~ $B^{1.86}$. Thus, we conclude that a strong temperature dependence of charge carrier concentrations is likely a key factor leading to the violation of Kohler's rule in MR behavior.

*3.1.3 Quantum oscillations and Berry phase*

Another characteristic feature of the MR(B) dependence is the appearance of SdH oscillations above 4 T, which are most clearly observed in the ΔMR(1/B) dependence obtained by subtracting a smooth polynomial background (Fig. 6(a)). Notably, the SdH oscillations exhibit beating-like behavior, indicating the presence of at least two closely spaced oscillation frequencies. To determine the oscillation frequencies, we performed Fast Fourier Transform (FFT) analysis with a Hanning window. Figure 6(b) clearly reveals three distinct oscillation frequencies: 86 T, 132 T, and 148 T. Since the cross-sectional area of the Fermi surface $A_F$ is related to the oscillation frequency through F = $\hbar A_F/2\pi e$, these oscillations correspond to different types of charge carriers associated with one hole ($h_1$) and two electron ($e_1$, $e_2$) Fermi surface pockets (see section 3.2). It should be noted that different studies of $WTe_2$ have reported between one to four fundamental oscillation frequencies, with significant dependence on crystal quality. For instance, Ref. [33] reported four frequencies (88 T, 133 T, 148 T, and 158 T), three of which closely match our results. The proximity of the two electron pocket frequencies ($e_1$ and $e_2$) is consistent with the observed beating pattern in the oscillations. We emphasize that the actual number of frequencies might be greater than the three observed in our study. As demonstrated by the two-band model analysis, different carrier types possess different mobilities, meaning the condition $\omega_c\tau \gg 1$ (or $\mu B \gg 1$) might only be satisfied at higher fields than those achieved in our experiments. This could explain why some potential oscillation frequencies remain unresolved in our measurements.

The effective mass for each carrier type observed in SdH oscillations can be determined within the Lifshitz-Kosevich (LK) formalism using the expression

$$A_{FFT} \sim \frac{\lambda T m^*/\langle B \rangle}{\sinh(\lambda T m^*/\langle B \rangle)}. \quad (5)$$

This expression describes the temperature dependence of the FFT amplitude with $m^*$ as the only fitting parameter, where $\lambda = (2\pi^2 k_B T)/\hbar e$ and $1/\langle B \rangle = (1/B_{min} + 1/B_{max})/2$ represents the average inverse field of the Fourier window [33, 34, 38]. In Fig. 6(c), the normalized FFT amplitudes for each observed pocket are shown as points, while the solid lines represent fits using equation (5). The resulting effective masses were determined to be $0.255m_0$, $0.265m_0$, and $0.285m_0$ for the $e_1$, $e_2$, and $h_1$ pockets, respectively. Notably, the reported values of $m^*$ in different studies show strong dependence on crystal quality and, consequently, on charge carrier mobility. For example, in Ref. [38] $WTe_2$ samples with RRR = 7.27 and carrier mobility of ~1300 cm$^2$/Vs yielded an effective mass of $m^* = 0.365m_0$ - significantly larger than our values obtained for higher-quality crystals with RRR = 38 and greater carrier mobilities (see Fig. 3). This comparison demonstrates how sample purity and electronic properties can substantially influence the measured effective masses in quantum oscillation experiments.

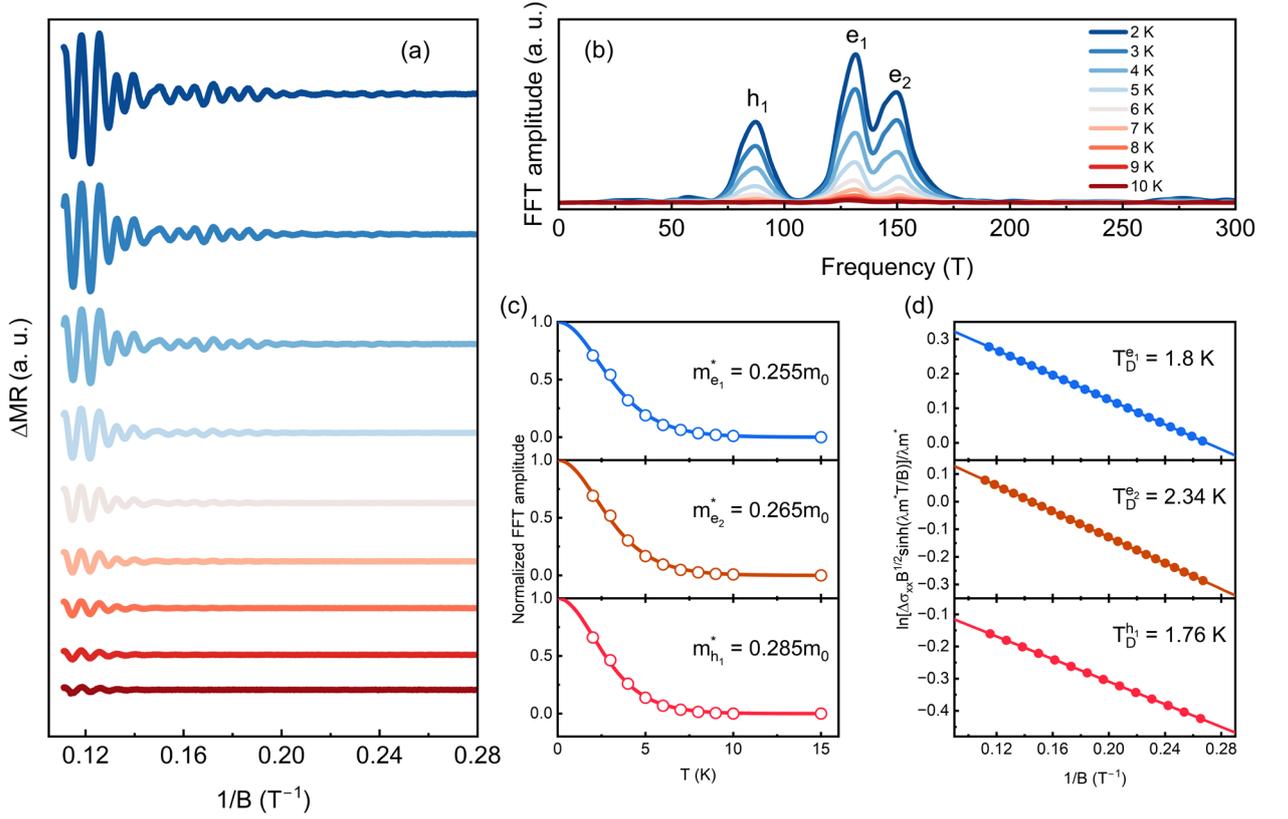

Fig. 6. (a) Oscillatory components of the magnetoresistivity ΔMR at different temperatures obtained after subtracting a smooth polynomial background. (b) Fast Fourier transform with a Hanning window, revealing three distinct oscillation frequencies. (c) Temperature dependence of the normalized FFT amplitudes. Solid lines denote fits based on LK formalism. (d) Dingle plot at T = 2 K.

A distinctive feature of topological materials involves the presence of Dirac or Weyl points representing linear band crossings. When an electron traverses a closed loop around such points in the Brillouin zone, it acquires an additional π Berry phase [16, 43]. In topological insulators, this π Berry phase arises exclusively for electron motion within the surface Brillouin zone (Fig. 7a). The situation becomes more complex in topological semimetals. Possible trajectories for a single Weyl point pair are shown in Fig. 7(b), including the characteristic Weyl orbit that connects surface Fermi arcs with bulk states. We emphasize that observation of Weyl orbits remains highly improbable in our experimental configuration since we investigate bulk crystal specimens.

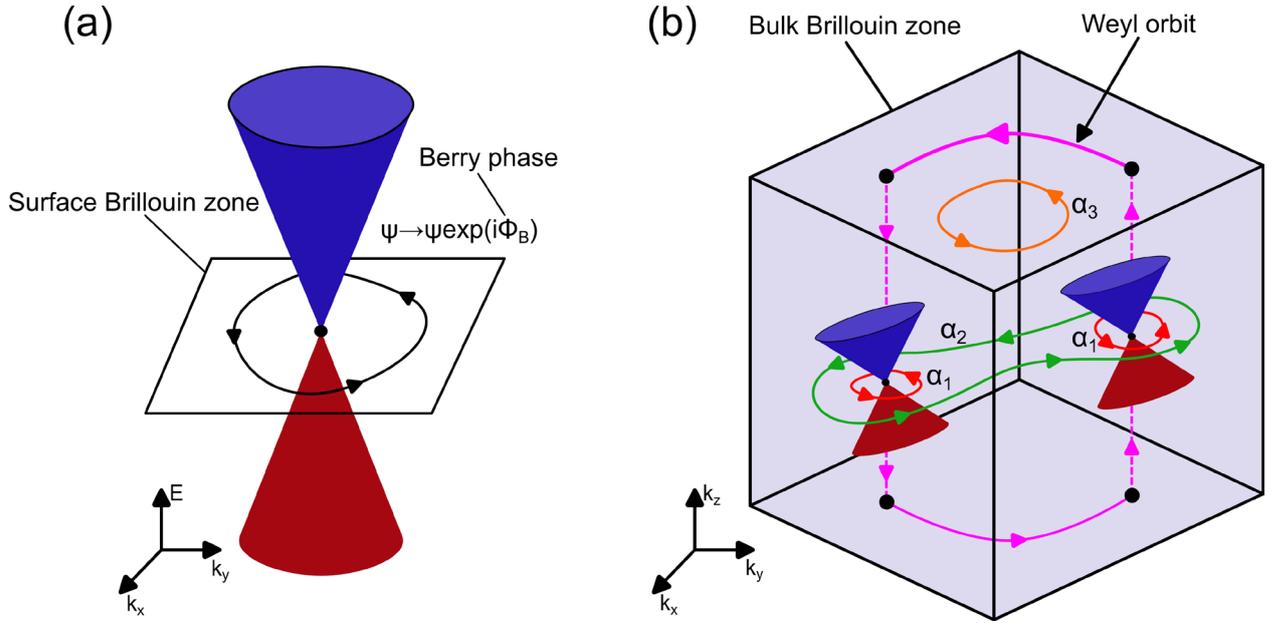

Fig. 7. Schematic illustration of possible electron trajectories in the Brillouin zone for (a) topological insulators and (b) Weyl semimetals that lead to the acquisition of a nontrivial Berry phase. In topological insulators, the Berry phase occurs exclusively in the surface Brillouin zone and is absent in the bulk. For topological semimetals where Weyl points reside in the bulk, there exist three types of trajectories in the bulk Brillouin zone that generate Berry phases: those enclosing a single Weyl point ($\alpha_1$-type), two or more Weyl points ($\alpha_2$-type), along with a special Weyl orbit trajectory connecting surface and bulk states. Additionally, numerous trivial trajectories ($\alpha_3$-type) exist where the Berry phase remains zero.

The standard approach for determining the Berry phase involves using the Lifshitz-Onsager quantization rule and constructing a Landau level (LL) fan diagram [16, 44]

$$n = \frac{F}{B} + \gamma, \qquad (6)$$

where $\gamma = -1/2 + \beta + \delta$ is the phase factor. Here, $\beta$ represents the Berry phase factor (0.5 for nontrivial $\pi$ Berry phase and 0 for zero Berry phase), while $\delta = \pm 1/8$ is the phase factor arising from the three-dimensional Fermi surface topology. To construct the LL fan diagrams, we assigned integer LL indices to the minima of the oscillatory component $\Delta\sigma_{xx}$, and half-integer indices to the maxima. Figures 8(a-c) present the LL fan diagrams at 2 K, 5 K, and 10 K. At 2 K and 5 K, the diagrams reveal three parallel lines with phase shifts, strongly indicating beating-like behavior in the oscillations. The phase shift points correspond precisely to the beating nodes, whose positions are marked by arrows in Fig. 8(d). Notably, the slope of these lines is 140 T, representing the average value of the two observed frequencies for the $e_1$ and $e_2$ pockets. At 10 K, only a single line remains due to thermal broadening of the Landau levels. This analysis demonstrates that the presence of multiple closely-spaced oscillation frequencies significantly complicates Berry phase determination using equation (6). As an alternative approach, we employed the multiband (LK) formalism [45-49], which accounts for

$$\Delta\sigma_{xx} = \sum_i A_i \sqrt{B} R_{T_i} R_{D_i} R_{S_i} \cos\left[2\pi\left(\frac{F_i}{B} + \gamma_i\right)\right]. \qquad (7)$$

Here $A_i$ represents a positive constant. $R_{T_i} = \frac{\lambda T m_i^*/B}{\sinh(\lambda T m_i^*/B)}$ describes the temperature dependence of oscillation amplitude damping. The factor $R_{D_i} = \exp(-\lambda T_{D_i} m_i^*/B)$ represents the Dingle factor, describing oscillation damping with varying magnetic field. Additionally, $R_{S_i} = \cos\left(\frac{\pi}{2} g_i \frac{m^*}{m_0}\right)$ denotes the spin factor arising from spin degeneracy lifting in magnetic field. The parameters $m_i^*$, $T_{D_i}$, $g_i$, $F_i$ и $\gamma_i$ represent effective mass, Dingle temperature, Landé factor, oscillation frequency, and phase factor respectively for different bands. The effective mass was fixed to values determined using equation (5). We found that this expression could best fit Δσxx in the case of i = 3 with frequencies very close to those determined from FFT, as expected (Fig. 8(b)). We also attempted to fit Δσxx using expression (7) with i > 3, however in this case we obtained negligibly small amplitude values for terms with frequencies differing from those determined from the Fourier transform, which also confirms that in the investigated field range we observe only three oscillation frequencies. The phase factors γ for pockets e₁ and e₂ were found to be 0.03 and 0.06 respectively, while the phase factor for h₁ equals 0.6. From this, one might assume that the trajectories in momentum space corresponding to e₁ and e₂ have α₁ or α₂ type, i.e., enclose a Weyl point, while trajectories for h₁ have α₃ type, i.e., do not contain a Weyl point (Fig. 7(b)). However, we would like to discuss that oscillation phase determination may not be unambiguous. It is clear that R_T and R_D are positive and cannot affect oscillation phase. However, it is worth noting that the LK formula also contains the factor R_S, which can influence the observed oscillation phase. This factor arises from spin degeneracy lifting in magnetic field, leading to two close oscillation frequencies and beating patterns. As a consequence, this leads to an additional factor $R_S = \cos\left(\pi \frac{E_S}{\Delta E}\right)$, where $E_S$ is the spin splitting energy and $\Delta E = \hbar\omega_c = \hbar eB/m^*$ is the Landau level spacing. If $E_S$ is not linear in magnetic field, $R_S$ may depend on magnetic field, leading to beating patterns. However, if spin splitting occurs due to Zeeman effect, then $E_S = g\mu_B B = gBe\hbar/2m_0$. Consequently, $R_S = \cos\left(\frac{\pi}{2} g \frac{m^*}{m_0}\right)$ becomes independent of magnetic field. In other words, the beating disappears and only one oscillation frequency remains. From this, it is clear that $R_S$ is also an oscillating term whose sign depends on both m* and g. Using equation (5), m* can be determined if the temperature dependence of oscillation amplitude is known, however the g-factor cannot be uniquely determined directly from standard transport measurements. It is worth noting that in most recent works determining Berry phase, the factor $R_S$ receives insufficient attention and is often not considered. However, this is particularly relevant for topological materials with strong spin-orbit and electron-electron interactions, whose g-factors can take various values ranging from 2 to 100 or higher. Moreover, the possible presence of non-trivial Berry phase does not allow us to establish what causes the observed oscillation phase. This can be easily understood since the case where $R_S > 0$ and $\beta$ = 0.5 is equivalent to the case where $R_S < 0$ and $\beta$ = 0.

Recently, the effect known as the spin-zero effect has been actively investigated, where the oscillation amplitude becomes zero at certain g-factor values due to $R_S = 0$. This effect is typically detected by studying magnetic field dependences of MR or spin torque at different angles between magnetic field direction and axis perpendicular to sample plane. Usually this effect is observed in two-dimensional systems [50-53], however recently it has also been discovered in kagome metal CsV$_3$Sb$_5$ [54] and topological semimetals ZrTe$_5$ [55], CaAgAs [56] and WTe$_2$ [57]. This effect occurs when the g-factor transitions from positive to negative $R_S$ values through zero or vice versa, indicating strongly anisotropic g-factor dependence on magnetic field direction and consequently strong Fermi surface anisotropy. As shown by authors in Ref. [57] the lowest estimate for WTe$_2$ yields g > 44.7, however determining the exact g-factor value for each carrier type is challenging, making it difficult to definitively determine the sign of $R_S$.

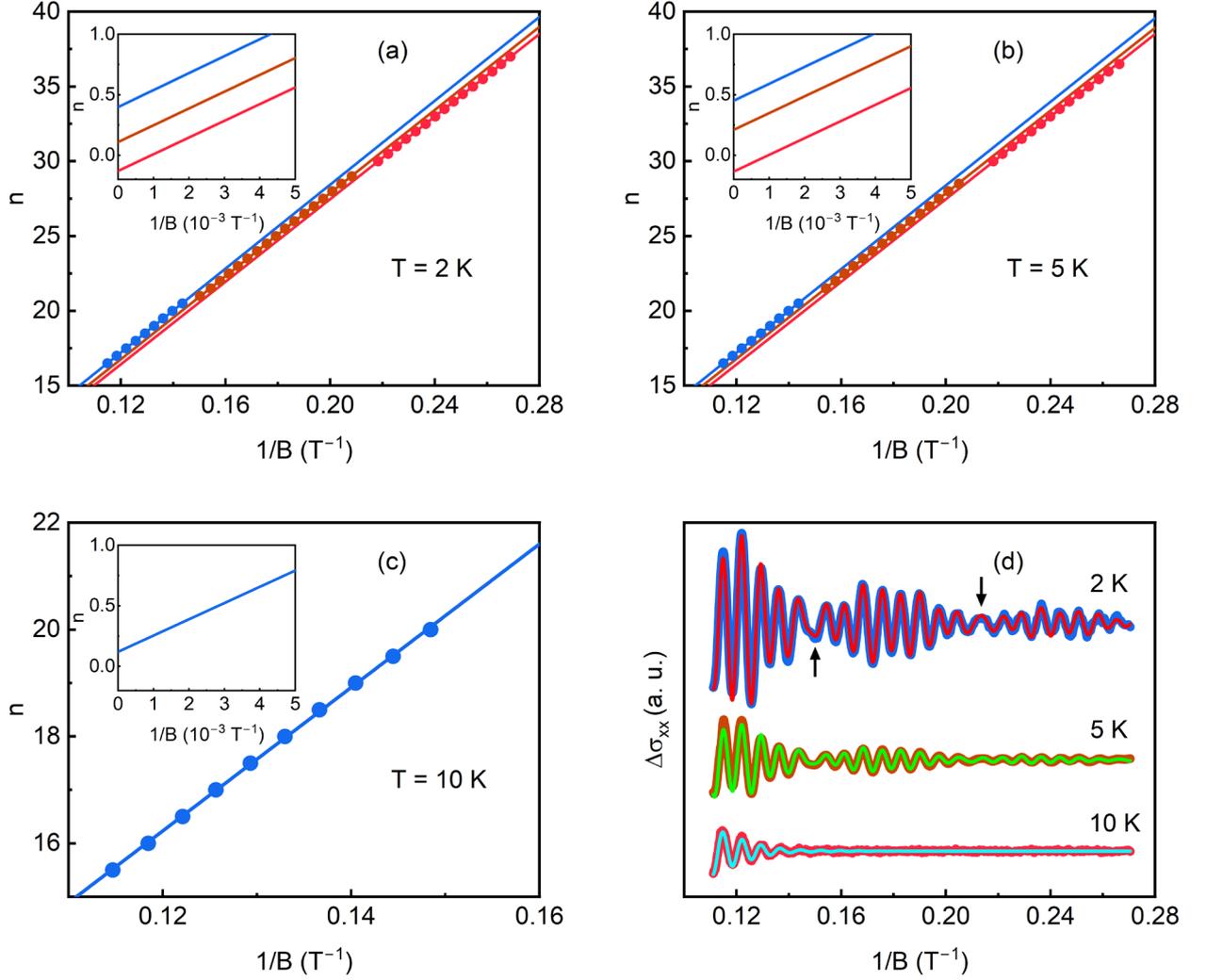

Fig. 8. (a) LL fan diagrams at 2, 5 and 10 K. At 2 and 5 K LL fan diagrams represents as three parallel phase-shifted lines, indicative of beating-like character of oscillations. (d) Oscillatory component of the magnetoconductivity Δσ$_{xx}$ at 2, 5 and 10 K. The red, green and cyan curve represents a fit using the LK formula with a multi-band model. Arrows mark the positions of beating nodes.

When determining oscillation phases, we implicitly assumed that $R_S > 0$, however it is clear that the opposite situation is possible, where the g-factor is such that $R_S < 0$, resulting in the observed oscillation phase being shifted by π. Furthermore, our calculations (see section 3.2) and ARPES research results [58] show that the Weyl point lies within the hole pocket rather than electron pocket, as we obtained from fitting expression (7) assuming $R_S > 0$. From this, we conclude that determining Berry phase solely from quantum oscillations can be incorrect and may lead to wrong conclusions, requiring combined techniques to determine carrier g-factors.

It should be noted that far from all works determine Berry phase in multi-band topological semimetals. In many studies, authors construct LL fan diagrams only for "major" oscillation extrema and found the corresponding Berry phase for one Fermi pocket [38, 59-62]. A similar approach, as in our work, was applied by authors in [63] studying quantum oscillations in TaP. They found that $\Delta\rho_{xx}$ could be successfully described by expression (7) with three oscillation frequencies, for two of which the Berry phase was close to π. At the same time, we were able to find only one work [38], that determined Berry phase in WTe$_2$. However, only one oscillation frequency was discovered in it. It is worth noting that in these works the authors also did not discuss the possible sign of $R_S$, which may cast doubt on conclusions about the actual topology of the band structure.

For determining the Dingle temperature, we constructed the Dingle plot $\ln\left(\Delta\sigma_{xx}B^{\frac{1}{2}}\sinh(\lambda m^*T/B)\right)/\lambda m^*$ for each term obtained using expression (7), corresponding to its respective Fermi surface sheet (Fig. 6(d)). In this expression $\Delta\sigma_{xx}$ represents magnetic field-dependent oscillation amplitude. T$_D$ is determined by fitting experimental data with a straight line, whose slope defines T$_D$. The Dingle temperature T$_D$ values were 1.8 K, 2.34 K, and 1.76 K for pockets e$_1$, e$_2$, and h$_1$ respectively. Similar values were obtained from fitting expression (3). We note that in most works studying SdH oscillations in WTe$_2$, Dingle temperature is not determined due to the multi-band nature of oscillations. Knowing m$^*$ и T$_D$, we can determine other electronic structure parameters for each Fermi surface pocket: quantum scattering time $\tau_q = \hbar/2\pi k_B T_D$, Fermi velocity $v_F = \hbar k_F/m^*$, Fermi energy $E_F = \hbar^2 k_F^2/2m^* = m^* v_F^2/2$, Dingle mean free path $l_D = v_F \tau_q$ and quantum mobility $\mu_q = e\tau_q/m^*$. The obtained parameters are presented in Table 1. Interestingly, $\mu_q < \mu_{Hall}$, which is not surprising since $\mu_q$ is determined by quantum scattering time $\tau_q$, which depends on all types of scattering, including small-angle scattering, while $\mu_{Hall}$ is determined by transport scattering time $\tau_{tr}$, which is mainly determined by large-angle scattering.

Table 1. Electronic structure and transport parameters determined from SdH oscillations analysis.

| Fermi pocket | $m^*/m_0$ | $T_D$, K | $k_F$, Å$^{-1}$ | $\tau_q$, 10$^{-13}$ s | $v_F$, 10$^7$ cm/s | $E_F$, meV | $l_D$, nm | $\mu_q$, 10$^3$ cm$^2$/Vs |
|---|---|---|---|---|---|---|---|---|
| e$_1$ | 0.255 | 1.8 | 0.063 | 6.75 | 2.87 | 60 | 194 | 4.65 |
| e$_2$ | 0.265 | 2.34 | 0.067 | 5.2 | 2.92 | 64 | 152 | 3.45 |
| h$_1$ | 0.285 | 1.76 | 0.051 | 6.9 | 2.06 | 34 | 142 | 4.26 |

Based on the obtained electronic structure parameters, we can schematically illustrate the electronic structure of WTe$_2$ along the Γ-X direction and the corresponding Fermi surface cross-sections in the k$_x$k$_y$ plane (Fig. 9). As shown by the two-band model analysis, WTe$_2$ is in a state close to compensation. Based also on ARPES experiments [58], we should expect the presence of another hole pocket h$_2$.

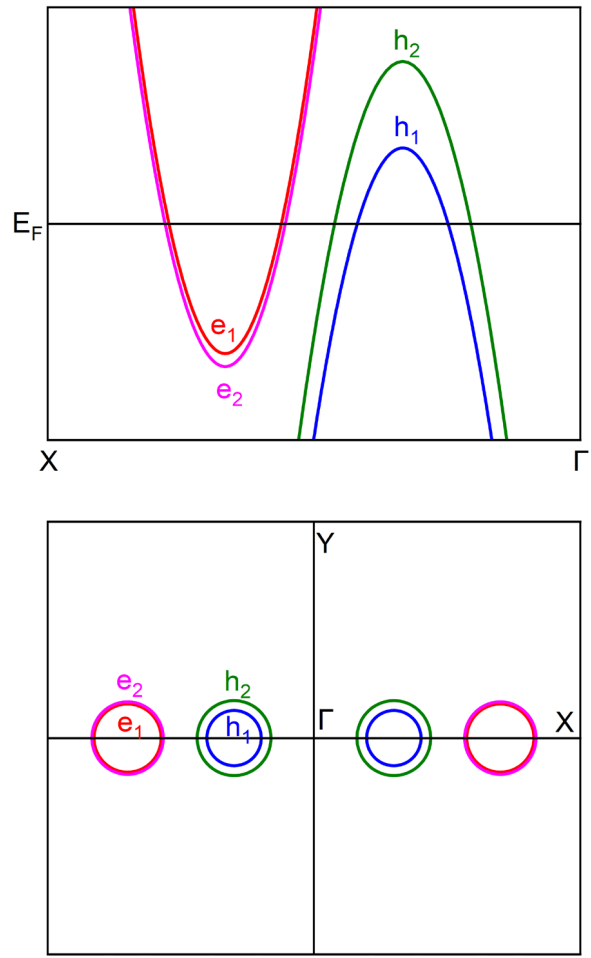

Fig. 9. Schematic representation of the band structure and Fermi surface of WTe$_2$ along the Γ-X direction, obtained from the SdH oscillations and two-band model analysis.

## 3.2 Electronic structure calculations

### 3.2.1 Band structure

It appears particularly pertinent to conduct a comparative analysis between our derived experimental data and the results obtained from first-principles calculations. In Fig. 10(a), the band structure of WTe$_2$ for DFT+U+SOC with U = 3 eV contains two small hole pockets near the Fermi level. One pocket located in the vicinity of the high symmetry point Γ and another one – along the direction Γ – X, these pockets have smaller versions of them at the same locations, created by band splitting due to spin-orbit coupling. Also these electron and hole pockets were identified in ARPES measurements [64]. The calculations show the presence of Weyl node within one of the hole pockets a little above the Fermi energy. There are also two larger almost degenerated electron pockets in the direction Γ – X. These pockets create Fermi surfaces detectable in the experiment and they also contribute to the small but noticeable density of states around the Fermi energy which is shown on the right of Fig. 10(a). Densities of states are plotted for the two prevailing electronic shells in this energy interval, namely, W 5d and Te 5p. These electronic states occupy both valence and conduction bands with the Te 5p states mostly filled.

The bands and densities of states of WTe$_2$ for DFT+U+SOC (U = 4 eV) mostly look similar to the previous case with U = 3 eV. The change in the U parameter results in the disappearance of the hole pocket in the vicinity of the high symmetry point Γ and also reduces the other two present electron and hole pockets. This results in a decrease in the densities of states at the Fermi energy, see Fig. 10(b). It was also noted that an increase of the U value results in the disappearance of the topological features in the band structure.

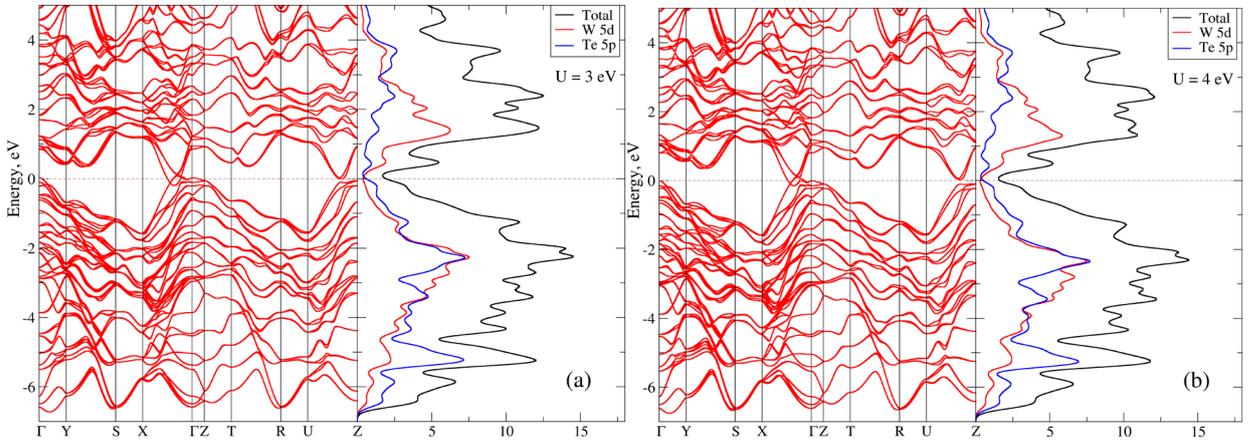

Fig. 10. Band structure and densities of states for WTe$_2$ calculated with DFT+U+SOC for the U value employed as (a) 3 and (b) 4 eV.

*3.2.2 Fermi surface*

Fermi surfaces of WTe$_2$ in Fig. 11(a) contain two pairs of sheets with surfaces within each pair having very close shape, volume and location in the Brillouin zone, which is the result of a band splitting due to spin-orbit coupling. The surfaces in the first pair correspond to the electron pockets in the band structure and the outer ones are shown in cyan in Fig. 11(a). The surfaces in the other pair correspond to the hole pockets and the outer ones are shown in magenta. These electron and hole pockets almost compensate for each other but still the electron ones are predominant.

In Fig. 11(b) one can see the sections of the Fermi surfaces with the largest surface area which is actually k$_z$ = 0 section of the Brillouin zone. The sections of the electronic Fermi surfaces have the largest surface area compared to the hole ones and the surface area of the one symmetric part of the section was calculated as 0.0346 Å$^{-2}$. In this figure, yellow squares indicate the positions of Weyl points located within the hole pockets. At the same time, from the analysis of oscillations using the multi-band LK formula (7) under the assumption that $R_S > 0$ for all pockets, we expect the presence of non-trivial Berry phase in the case of electron pockets. From this it follows that it is possible that the g-factor is such that $R_S < 0$, which would lead to a phase shift of oscillations by π. From this we want to emphasize once again that determining oscillation phase solely from the SdH effect can be ambiguous and for correct determination of Berry phase, one should know the g-factors of each carrier type.

The Fermi level was also shifted down by 40 meV in energy to mimic experimental to reduce the size of the electronic Fermi surfaces and the surface area of this section was calculated as 0.0177 Å$^{-2}$. This value is close to the experimental value 0.0141 Å$^{-2}$ for electronic pockets. Other studies report same number and similar shapes of the Fermi surfaces besides the absence of the elliptical one right around the high symmetry point Γ, the presence of which might be due to the small crystal structure differences of the studied samples [33, 64-66]. There are also reports that position of the Fermi level changes with temperature in the range of about 50 meV [65, 67], which can largely impact the size of the electron and hole pockets as it was shown in this work. On the contrary because of the energy shift, two pairs of the hole pockets became much larger and connected creating two big hole pockets, that form two large Fermi surfaces.

In Fig. 11(c), one can see that the change in the parameter U lowers the size of every present Fermi surface with the hole surface around the Γ point completely disappearing. The surface area of the one symmetric part of the electronic Fermi surface section is equal to 0.0245 Å$^{-2}$. Overall this picture very much resembles the schematic representation of Fermi surfaces in k$_x$k$_y$ plane of the Brillouin zone in Fig. 9.

By comparing our results with the published ARPES measurements we find that U = 3 eV, Fig 11(b), has better correspondence with the experiment showing all of the identified hole and electron pockets [68] and the presence of Weyl points. Also, this value was obtained in our additional calculations for the experimental crystal structure parameters reported in this work, see Section 2. Materials and methods. The Hubbard parameter value U = 4 eV better reflects the sizes of the Fermi surfaces compared to data from Shubnikov-de Haas oscillations, but in this case, we do not observe Weyl nodes and topology around the Fermi level is trivial. Other works also report non-trivial band structure topology [58] and non-trivial Berry phase was extracted from quantum oscillations. Thus, the calculated value U = 3 eV, Fig 11(b), better correlates with the published experimental studies and even though in this case we observe additional unidentified from quantum oscillations hole pocket around point Γ, it is actually quite small and very sensitive to crystal structure parameters.

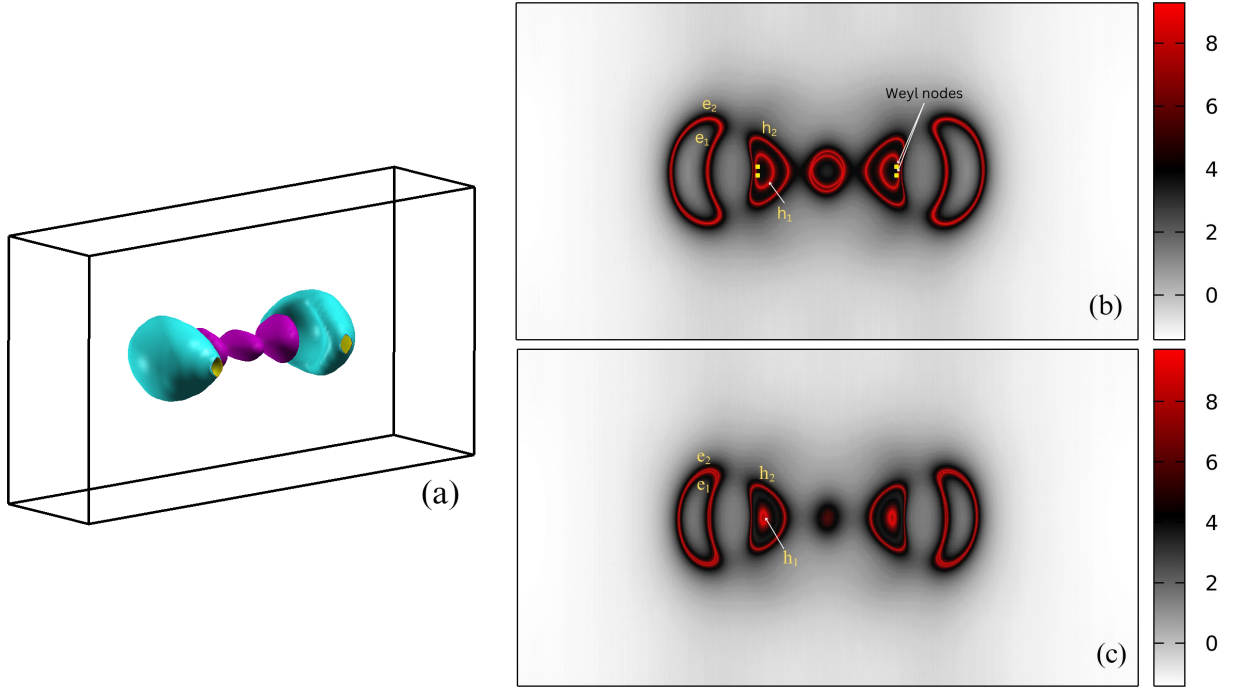

Fig. 11. (a) 3D Fermi surfaces for WTe$_2$ calculated within DFT+U+SOC, U = 3 eV. (b) k$_z$ = 0 plane in the Brillouin zone showing sections of the Fermi surfaces calculated with U = 3 eV. Yellow squares show the calculated positions of the Weyl nodes. (c) k$_z$ = 0 plane in the Brillouin zone showing sections of the Fermi surfaces calculated with U = 4 eV.

## 4. Conclusion

In conclusion, we have conducted a comprehensive investigation of the electronic structure and transport properties of the topological semimetal WTe$_2$. The temperature dependence of the electrical resistivity demonstrates metallic behavior throughout the entire temperature range, with ρ ~ T$^2$ at temperatures up to 70 K, which is associated with both electron-electron scattering and the interference mechanism of "electron-phonon-surface" scattering. At the same time, in a magnetic field of 9 T, the minimum appears in the ρ(T) dependence, caused by the transition from effectively high to effectively weak magnetic fields. Analysis of the field dependencies of σ$_{xx}$(B) и σ$_{xy}$(B) showed that WTe$_2$ is in a state close to electron-hole compensation with high values of MR ~1900% and carrier mobilities, which emphasizes the high quality of the studied sample. In addition, it was found that MR in WTe$_2$ does not follow the classical Kohler's rule MR = f(B/ρ$_0$), but can be well described by the extended Kohler's rule MR = f(B/n$_T$ρ$_0$) with the parameter n$_T$ reflecting the temperature dependence of carrier concentration, which indicates the complex nature of carrier scattering and temperature dependence of their concentration. Analysis of the SdH oscillations allowed us to identify three characteristic frequencies corresponding to different pockets on the Fermi surface. The obtained electronic structure parameters, including effective masses of carriers and Dingle temperatures, confirmed the high quality of the investigated single crystal. Special attention was paid to the issue of determining the Berry phase, where it was shown that traditional analysis methods can give ambiguous results due to the possible influence of the spin factor and g-factor anisotropy. The results of electronic structure calculations within the DFT+U+SOC approach are in good agreement with experimental data and demonstrate the sensitivity of the band structure to the value of the Hubbard parameter. The presence of Weyl points in the calculated band structure and their location within the hole pockets is of particular interest for understanding the potential topological properties of $WTe_2$. The obtained results are important for the potential application of this material in spintronic devices and other modern technologies. For final confirmation of the topological nature of the observed effects, further research is needed, including direct measurement of carrier g-factors.

**Acknowledgements**

The study of Kohler's rule, quantum oscillations and Fermi surface in $WTe_2$ (sections 3.1.2, 3.1.3, 3.2.2) were carried out within the framework of the state assignment of the Ministry of Science and Higher Education of the Russian Federation for the IMP UB RAS and Ural Federal University (project within the Priority-2030 Program). Synthesis of $WTe_2$ single crystal, investigation of magnetotransport properties and band structure calculations (sections 2, 3.1.1, 3.2.1) were supported by the Russian Science Foundation (project No. 24-72-00168, https://rscf.ru/en/project/24-72-00168/, M.N. Mikheev Institute of Metal Physics, Ural Branch of the Russian Academy of Sciences, Sverdlovsk Region).